# Deploying Wireless Networks with Beeps

Alejandro Cornejo
MIT
acornejo@csail.mit.edu

Fabian Kuhn
University of Lugano
fabian.kuhn@usi.ch

February 17, 2010

**Abstract**

We present the *discrete beeping* communication model, which assumes nodes have minimal knowledge about their environment and severely limited communication capabilities. Specifically, nodes have no information regarding the local or global structure of the network, don't have access to synchronized clocks and are woken up by an adversary. Moreover, instead on communicating through messages they rely solely on carrier sensing to exchange information. This model is interesting from a practical point of view, because it is possible to implement it (or emulate it) even in extremely restricted radio network environments. From a theory point of view, it shows that complex problems (such as vertex coloring) can be solved efficiently even without strong assumptions on properties of the communication model.

We study the problem of *interval coloring*, a variant of vertex coloring specially suited for the studied beeping model. Given a set of resources, the goal of interval coloring is to assign every node a large contiguous fraction of the resources, such that neighboring nodes share no resources.

To highlight the importance of the discreteness of the model, we contrast it against a continuous variant described in [17]. We present an $\mathcal{O}(1)$ time algorithm that terminates with probability 1 and assigns an interval of size $\Omega(T/\Delta)$ that repeats every $T$ time units to every node of the network. This improves an $\mathcal{O}(\log n)$ time algorithm with the same guarantees presented in [17], and accentuates the unrealistic assumptions of the continuous model. Under the more realistic discrete model, we present a Las Vegas algorithm that solves $\Omega(T/\Delta)$-interval coloring in $\mathcal{O}(\log n)$ time with high probability and describe how to adapt the algorithm for dynamic networks where nodes may join or leave. For constant degree graphs we prove a lower bound of $\Omega(\log n)$ on the time required to solve interval coloring for this model against randomized algorithms. This lower bound implies that our algorithm is asymptotically optimal for constant degree graphs.

# 1 Introduction

Communication models face the unavoidable tension between their practicality and their potential for designing interesting yet provably correct algorithms. With enough assumptions concerning the knowledge of the environment and the communication capabilities of nodes, it is not difficult to design efficient and elegant distributed algorithms. However, it is often difficult (if not impossible) to translate these algorithms to the real world. On the other hand, communication models which are cluttered with physical details encumber designing algorithms, and makes it significantly more complicated to prove correctness or efficiency.

This motivates the study of models such as the *discrete beeping* model considered in the present paper. The model makes little demands on the communication devices, nodes need only be able to do *carrier sensing* and differentiate between silence and the presence of a jamming signal. Carrier-sensing can typically be done much more reliably and requires significantly less energy and other resources than transmitting and receiving actual messages , see e.g. [7]. Besides requiring reliable carrier sensing, we make almost no assumptions. In particular, we do not assume knowledge of the local or global structure of the network or synchronized clocks. Further, we assume that an adversary controls when processors are woken up.

We show that even such a "weak" model allows for interesting algorithms for non-trivial tasks. In particular we focus on the problem of *interval coloring*, a variant of classic vertex coloring. Given a set of resources, the goal of interval coloring is to assign each node a large subset of contiguous resources such that neighboring nodes obtain disjoint resources. Similar to vertex coloring, interval coloring is a useful building block to establish a reliable Medium Access Layer (MAC), as it can be used to e.g. compute time or frequency division multiple access (TDMA or FDMA) schedules that avoid conflict between potentially interfering nodes. In some sense, interval coloring is even better suited for these tasks than standard graph coloring. While in a standard coloring, every node gets assigned a single color (a single slot or frequency), in an interval coloring, we can assign larger intervals to certain nodes (e.g. to nodes with a small degrees). An interval then corresponds to multiple consecutive colors in a standard coloring context.

Moreover, by relying exclusively on carrier sensing, the beeping model becomes specially well-suited for coordination tasks in wireless networks for various reasons, for example: ⋄ Most prior work [1, 3, 4, 9, 11, 14, 18, 23, 25] on coloring assumes some existing infrastructure to reliably exchange messages. If used as a building block to e.g. compute a TDMA schedule, these algorithms suffer from a chicken-and-egg problem; such colorings cannot be computed without a reliable MAC layer, however to achieve a reliable MAC layer one first needs to compute a coloring. ⋄ The presence of a signal can be reliably detected by carrier sensing at lower receiving power than would be required to correctly decode a message. Hence, carrier sensing can be used to communicate more energy efficiently and over larger distances than when transmitting regular messages. For example, by default the NS2[26] simulator uses a carrier sensing range that is more than twice as large as the transmission range. Therefore, the beeping model (carrier sensing) can directly be used to compute a 2-hop interval coloring of the communication graph (for regular transmission), a necessity when using the coloring for a MAC layer that avoids hidden terminal collisions. ⋄ Although IEEE 802.11 and Bluetooth share the same frequency spectrum, they use incompatible modulation and encoding schemes. However since carrier sensing only detects the presence of a signal, it is potentially possible for a IEEE 802.11 radio to detect the presence of a Bluetooth jamming signal and vice versa. Therefore, algorithms for the beeping model could be used to allow these two seemingly incompatible devices to agree on a non-conflict transmission schedule thereby allowing them to coexist in a non-destructive fashion.

**Contributions**

We assume that there is a common globally known period length $T$. This is a parameter of the algorithms which captures the number of resources to be shared (e.g. the number of available frequencies in FDMA). The paper has three main contributions.

First, we significantly improve a result from [17] for a continuous variant of the beeping model. In [17],

it is shown that in $\mathcal{O}(\log n)$ periods, it is possible to assign an interval of length $\Omega(T/d^{\max}(v))$ to each node $v$, where $d^{\max}(v)$ is the largest degree in the 1-neighborhood of $v$. We describe a simpler algorithm that improves the results of [17] by computing an interval coloring with the same properties in a constant number of periods. Our result highlights the unrealistic assumptions behind the continuous model.

Second, we give a discrete variant of the beeping model and describe a Las Vegas randomized interval coloring algorithm for the discrete model. The algorithm computes intervals of length $\Omega(T/d^{\max}(v))$ in $\mathcal{O}(\log n)$ periods with probability $1 - \frac{1}{n}$. Furthermore, we describe how to adapt the algorithm to work in a dynamic graph setting where nodes can join and leave arbitrarily. A new node obtains an interval at most $\mathcal{O}(\log n)$ periods after joining the network, and a node only recomputes its interval if the size of its neighborhood becomes drastically smaller. The correctness proof of both the static and dynamic versions of the algorithm rely on a balls and bins analysis which, due to lack of space, is presented in Appendix A.

Finally, for a local broadcast model with constant size messages, we prove a lower bound of $\Omega(\log n)$ time against randomized algorithms that solve $\mathcal{O}(\Delta)$ vertex coloring (or interval coloring with intervals of size $\Omega(T/\Delta)$). For the discrete beeping model this implies a lower bound of $\Omega(\log n)$ periods for constant-degree graphs and $\Omega(\log n/\Delta)$ for general graphs. Moreover, if we restrict the number of beeps per period to $\mathcal{O}(1)$ it yields a lower bound of $\Omega(\log n / \log \Delta)$ for general graphs.

**Related Work**

Using carrier sensing for distributed computation is not novel. Scheideler et al. [21] considered a model where in addition to sending and receiving messages, nodes can perform physical carrier sensing, and described how to approximate the minimum dominating set problem under this model. Flury and Wattenhofer [7] demonstrate how to use carrier sensing as an elegant and efficient way for coordination in practice.

Our beeping model is a discretized variant of the desynchronization model first introduced by [6]. Degesys et al. [6] considered only complete graphs, and proved the eventual convergence of a biologically inspired algorithm DESYNC to a *'desynchronized state'* and conjectured a running time of $\mathcal{O}(n^2)$. Degesys and Nagpal [5] experimentally studied the performance of DESYNC in multi-hop topologies. They proved that a desynchronized state exists for 2-colorable graphs and Hamiltonian graphs, and posed the open problem of proving that a desynchronized state exists for all graphs. Later Motskin et al. [17] studied interval coloring under the same desynchronization model. In addition to assuming the continuous variant of the model, [17] assumes that nodes have knowledge of their own degree and that they are able to exchange this information to compute the maximum neighbor degree over their 1-hop neighbors. It is not clear how nodes should obtain the maximum degree among their neighbors without reliably transmitting messages (in which case we do not need to "zurueckgreifen" to the beeping model). Further, as we show in Section 4, their assumptions are too strong and allow for constant time solutions. This in particular motivates the strictly weaker discrete beeping model.

Coloring the nodes of a graph is one of the most fundamental combinatorial optimization problems in computer science and has therefore been widely studied, also in a distributed context. The work on distributed coloring algorithms started with the seminal work of Linial [14] and includes a large number of papers (see e.g. [1, 3, 4, 9, 11, 13, 18, 23, 25]). The best bounds are known for randomized algorithm and they are $\mathcal{O}(\sqrt{\log n} + \log \Delta)$ for $(\Delta+1)$-colorings (i.e., the number of colors needed by the sequential greedy algorithm) and $\mathcal{O}(\sqrt{\log n})$ for $\mathcal{O}(\Delta)$-colorings [11, 25]. Interesting in the context of TDMA schemes for wireless networks might be [12] where it is shown how to compute a coloring where each node with degree $d$ obtains an $\Omega(1/d)$-fraction of the colors in a single communication round (i.e., nodes just need to learn the identifiers of all neighbors). Coloring in unstructured radio networks (with collisions) was considered by [16], where a randomized algorithm to compute $\mathcal{O}(\Delta)$ colorings in $\mathcal{O}(\Delta \log n)$ rounds is described (later improved in [24] to $\mathcal{O}(\Delta + \log \Delta \log n)$ rounds). In addition to the theoretical work on distributed coloring, there are many papers that describe some variant of coloring in order to compute TDMA schedules or similar MAC schemes (see e.g. [2, 8, 10, 15, 19, 20, 27]).

## 2 Model and Definitions

We consider a wireless network model that is as primitive as possible. In contrast to standard communication models, nodes cannot exchange messages reliably (message passing) or unreliably (unstructured radio networks), instead nodes rely entirely on carrier sensing. At any particular time, a node can be in beeping or listening mode. When a node is listening, it can only distinguish between silence or the presence of one or more beeps. This model is weaker than collision detection since nodes cannot distinguish between a single beep and a collision of two or more beeps. Moreover, a beep conveys less information than a bit, and although one could conceive coding schemes to encode bit messages using beeps, this would require additional overhead and be susceptible to collisions, thus we focus on different techniques.

The communication network is modeled as an undirected graph $G = (V, E)$, $|V| = n$, where the set $V$ of nodes of $G$ represents the set of wireless devices. There is an edge $\{u, v\} \in E$ if and only if $u$ can listen to a beep emitted by $v$ and viceversa. For a node $u \in V$, let $N(u) := \{v \in V \mid \{u, v\} \in E\}$ be the set of neighbors of $u$, and let $d(u) = |N(u)|$ be its degree. A *phase* refers to a time point (in the continuous model) or a time slot (in the discrete model) measured relative to the beginning of the last period. We will use phases to capture the time at which different beeps are heard with respect to the local clock of each node. Given a set $S$ of phases, we define $S[a, b]$ to be the subset of phases in the range $[a, b]$ in $S$. To correctly account for ranges that cross the period boundary, we give a formal definition. Let $\tau$ be the period length (in the continuos model the period length is $T$ time units, while in the discrete model the period length is $Q$ time slots), and let $x = a \mod \tau$ and $y = b \mod \tau$. If $x \leq y$, $S[a, b] = \{p \in S \mid x \leq p \leq y\}$, otherwise $S[a, b] = \{p \in S \mid p \leq x \vee y \leq p\}$.

Consider neighboring nodes $u$ and $v$, suppose that node $u$ executes some event $e_u$ at local time $t_u$ which is instantaneously observed by node $v$ at local time $t_v$. If $u$ and $v$ measure time using unsynchronized local clocks, in general $t_u \neq t_v$. If $t_u$ represents the time of occurrence of some event with respect to node $u$ we use $\mathring{t}_u$ to represent the time of occurrence of the event in a global reference frame, hence in the previous example $\mathring{t}_u = \mathring{t}_v$.

We say that an event happens almost surely if it happens with probability one, an event happens with high probability if it occurs with probability at least $1 - \frac{1}{n}$. Let $\mathfrak{U}(a, b)$ denote the continuous uniform distribution in the range $[a, b]$ and $\mathfrak{U}[a..b]$ denote the discrete uniform distribution in the range $[a..b]$.

We assume that nodes wake up asynchronously and the wake-up pattern is determined by an adversary. Upon waking up, a node does not know anything about the topology of the graph, an estimate of the network size $n$ or the maximum degree of the graph $\Delta$. Similarly, nodes do not know their neighbor set or have an estimate of the size of this set. Furthermore, nodes do not have unique identifiers and the structure of the communication graph $G$ is not restricted in any way (e.g. by requiring $G$ to be a unit disk graph, a bounded independence graph, or any other special type of graph considered in the wireless networks literature [22]). Every node has access to a local clock, where the local clock of every node advances at the same rate and has no drift, however we do not assume clocks to be synchronized.

We believe that the above model is simple enough to enable algorithms designed for this model to be implemented and executed in real hardware, and yet complex enough to allow for the design of interesting algorithms with strong theoretical guarantees. We consider two variants of the basic model, a continuous version and a discrete version.

**Discrete Model**

Time is divided into slots of length $\mu$, where $\mu$ depends on the physical characteristics of the wireless devices and of the communication medium. There is a known integer $Q > 0$ that denotes the number of slots per period, and is related to the number of resources available. Hence, the period length is $T = Q\mu$. Although we do not assume synchronized clocks, we assume that slots boundaries are synchronized, i.e., all nodes start new slots at the same time. Note that at the cost of small constant factors and more technical arguments, all results obtained in this paper can also be achieved in a model with unsynchronized slot boundaries.

In each slot $s$, each node $v$ can either listen or beep for the whole duration of $s$. If a beep is emitted by node $u$ at slot $s$, it is heard by any neighboring node $v \in N(u)$ that is in listening mode in slot $s$. In particular the operation listen$[m]$ puts the node in listening mode for the next $m$ slots and returns the set of slots where it detected a beep. The operation beep emits a beep for the duration of the current slot, and returns no feedback.

**Continuous Model**

All nodes share some period length $T$ and a beep can be infinitely short (i.e., a unit impulse function). If a beep is emitted by node $u$ at time $t$, it is heard by any neighboring node $v \in N(u)$ that is in listening mode at time $t$. In particular the operation listen$(\delta)$ puts the node in listening mode for the next $\delta$ units of time and returns the set of time points where it detected beeps. The operation beep emits an infinitely short beep and returns no feedback. We will discuss some of the shortcomings of this variant in Section 4.

## 3 Interval Coloring

One of the central motivations behind vertex coloring in distributed environments is to use it as a building block for MAC protocols. In this setting the number of colors used translates to the number of communication channels used, and thus fewer colors imply higher throughput. In general we are interested in efficient (polylog or better) algorithms that produce vertex colorings with $\mathcal{O}(\Delta)$ colors, where $\Delta$ is the maximum degree. However, most known distributed algorithms for coloring are based on the assumption that there is already an infrastructure to reliably transmit messages with neighboring nodes, which makes them unsuitable for MAC protocols.

The input of interval coloring is a set of resources, and the output assigns to each node a contiguous subset of the resources, where the resources of neighboring nodes are disjoint. For example, in the continuous beeping model, interval coloring outputs at each node $v$ a tuple $\langle p_v, I_v \rangle$, where $p_v$ is the phase and $I_v$ is the interval length, such that for every pair of neighbors $\{u, v\} \in E$ the intervals $[\mathring{p}_v - I_v, \mathring{p}_v]$ and $[\mathring{p}_u - I_u, \mathring{p}_u]$ are disjoint. Analogous to $\mathcal{O}(\Delta)$-vertex colorings, we are interested in $\Omega(T/\Delta)$-interval colorings, where the smallest interval length is $\Omega(T/\Delta)$.

*Hardness of Interval Coloring.* Discrete interval coloring is strongly related to vertex coloring. Consider a fixed node $v$ and let $p_v$ be the phase it computed when solving interval coloring and $\Theta_v$ be its clock offset with respect to some arbitrary but absolute reference frame. Define the color of $v$ as $c_v = p_v + \Theta_v \pmod{Q}$. Observe that this defines a valid vertex coloring with $\mathcal{O}(\Delta)$ colors, since $Q \in \mathcal{O}(\Delta)$ and by definition of interval coloring for any two neighbors $u$ and $v$ it follows that $p_v + \Theta_v \neq p_u + \Theta_u$ and thus $c_v \neq c_u$. Therefore even in executions where all nodes have either synchronized clocks or wakeup at the same time, a $\Omega(T/\Delta)$-interval coloring is at least as hard as $\mathcal{O}(\Delta)$-vertex coloring.

## 4 Continuous Interval Coloring

We essentially use the same model as Motskin et al. in [17], and adhering to it we also assume each node $v$ knows its own degree $d(v)$ and the maximum degree of its 1-hop neighbors $d^{\max}(v)$. Motskin et al. [17] described a randomized algorithm that solves continuous interval coloring and terminates with high probability in a logarithmic number of periods. In contrast, we present a randomized algorithm that solves the same problem but terminates almost surely in a constant number of periods. While describing the algorithm we expose the flaws of this model that make such an algorithm possible.

*Algorithm Description.* Since nodes can emit an infinitely short beep at any point in time, then if two nodes choose to beep at random times in the interval $[0, T]$, their beeps collide with probability zero. We will exploit this property with the greedy algorithm BEEPFIRST. Informally speaking, the BEEPFIRST algorithm searches for the first available time where a node can beep while respecting a buffer of size $b_v$ around existing beeps. To ensure that no two nodes choose the same time to beep, the buffer size is randomized with a continuous variable.

More precisely, the algorithm has a parameter $\varepsilon \in (0,1)$ which affects the size of the resulting intervals. In the initialization state each node $v$ sets its interval length to $I_v = (1-\varepsilon)T/2(d^{\max}(v)+1)$ and chooses $\varepsilon_v \in \mathfrak{U}[0,\varepsilon]$ to randomize its start time and set its buffer length to $b_v = (1-\varepsilon_v)T/2(d(v)+1)$.

In the searching state, nodes listen for one full time period $T$ recording the phases at which beeps were heard. If a node hears no beeps in this period it sets $p_v = 0$ and goes to the stable state. Otherwise nodes search for the first phase $p_v$ such that i) in the previous period no other node beeped in the interval $[p_v - b_v, p_v + b_v]$, and ii) in this period no other node beeps on the interval $[p_v - b_v, p_v]$. Once such a phase is found, nodes beep to reserve it and listen for whatever remains of the period, switching to the stable state. Once a node becomes stable, it remains stable thereafter, beeping at the same phase every period.

---

**Algorithm 1** BEEPFIRST running at node $v$

1: $\varepsilon_v \leftarrow \mathfrak{U}(0, \varepsilon)$  ▷ Initialize
2: $I_v \leftarrow (1-\varepsilon)\frac{T}{2(d^{\max}(v)+1)}$, $b_v \leftarrow (1-\varepsilon_v)\frac{T}{2(d(v)+1)}$
3: $\texttt{listen}(\varepsilon_v)$ (* randomized start time *)
4: $S \leftarrow \texttt{listen}(T)$  ▷ Search
5: $p_v \leftarrow 0$
6: **while** $\exists$ beep in $S[p_v - b_v, p_v + b_v]$ **do**
7: $\quad t_v \leftarrow p_v$
8: $\quad p_v \leftarrow b_v +$ time of last beep in $S[p_v - b_v, p_v + b_v]$
9: $\quad S \leftarrow S \cup \texttt{listen}(p_v - t_v)$
10: **end while**
11: $\texttt{beep}, \texttt{listen}(T - p_v)$  ▷ Stable
12: **loop**
13: $\quad \texttt{listen}(p_v), \texttt{beep}, \texttt{listen}(T - p_v)$
14: **end loop**

---

Fix a node $v$ in the searching state, and observe that the separation between any beeps node $v$ hears, is at most $2b_v$ (otherwise it would have exited the search state). Assume in a period node $v$ hears at most one beep from each neighbor (a slightly more technical argument shown in Appendix B proves the same result without this assumption). Therefore node $v$ hears at most $d(v)$ beeps in one period, which means that after time $d(v)2b_v < T$ in the searching state node $v$ finds a proper phase to beep and enters the stable state.

**Lemma 4.1.** *The searching state of* BEEPFIRST *lasts less than one period.*

By construction node $v$ will select $p_v = 0$ or $p_v = p_u + b_v$ where $p_u$ is the phase of node $u$. However, both the starting time and the buffer length are randomized using a continuous probability distribution. Therefore, with probability one no two nodes will ever select the same phase. (The same argument is used by Motskin et al. [17] to prove that neighbors "pick the *exact* same start time with probability 0".) Which is captured by the following proposition.

**Proposition 4.2.** *Given a pair of nodes $u$ and $v$ (where $u \neq v$) at any point during the execution of* BEEPFIRST *almost surely $\mathring{p}_u \neq \mathring{p}_v$.*

From Proposition 4.2 it follows that given two neighboring nodes which have selected phases, one selected an earlier phase than the other, and therefore by construction the intervals output by BEEPFIRST do not overlap (proved in Appendix B).

**Lemma 4.3.** *Let $u$ and $v$ be two neighboring nodes in a stable state of BEEPFIRST, then their intervals do not overlap ($\mathring{p}_u \notin [\mathring{p}_v - I_v, \mathring{p}_v + I_v]$).*

Finally, Lemmas 4.1 and 4.3 imply the following theorem.

**Theorem 4.4.** *The continuous interval coloring algorithm computes an $\Omega(T/\Delta)$-interval coloring and terminates almost surely in $O(1)$ time.*

If instead of setting the interval length in the initialization phase we delayed it until the stable phase by setting it to the largest value such that $[p_v - I_v, p_v + I_v]$ does not contain any beeps, we would get a slightly stronger result which does not require knowledge of $d^{\max}(v)$. This hints at two flaws in this model i) It assumes knowledge of $d(v)$ and $d^{\max}(v)$, where neither is trivial to compute. ii) The algorithm's correctness relies on computation with arbitrary real numbers and sampling from continuous probability distributions.

## 5 Discrete Interval Coloring

We now turn our attention to a more realistic model where beeps occur at discrete times and have a minimum length; thus the probability distributions involved are discrete and finite. We present a Las Vegas randomized algorithm for $\Omega(Q/\Delta)$-interval coloring that terminates with high probability in $\mathcal{O}(\log n)$ periods. This requires $Q \geq \Delta$ and in particular we assume $Q = \kappa \Delta$ where $\kappa$ is a large enough constant ($\kappa \geq 3/\eta$ suffices, for $\eta$ to be fixed later).

*Algorithm Description.* The JITTERANDJUMP algorithm relies on three key insights: i) The number of beeps heard by a node is a good estimate of its degree. ii) By adding a small random jitter to every beep, neighboring nodes which beep at the same slot can detect the collision with constant probability. iii) If a node jumps into a random slot which is surrounded by "enough" empty slots it finds a non-conflicting interval assignment with constant probability.

Specifically, all nodes executing JITTERANDJUMP are initially uncolored. Nodes become colored as soon as they believe to have found their interval. Except for the first period (where nodes listen without beeping), all nodes beep once per period. Since nodes beep at most once per period, then in a single period a node can hear at most two beeps per neighbor. Hence if $\tilde{d}_v$ is the number of beeps observed by node $v$ during a period, then $1 \leq \tilde{d}_v \leq 2d(v)$.

To resolve collisions, if node $v$ has decided to beep at the slot $p_v$, it chooses choses at random $jitter_v \in \mathfrak{U}[0..1]$, and beeps at $p_v + jitter_v$ instead. If a colored node detects a beep one slot before, or two slots after its own beep, it becomes uncolored.

Each node $v$ sets the buffer length $b_v = \eta \frac{Q}{\tilde{d}_v + 1}$ to a fraction of the period proportional to its degree estimate, where $\eta$ is a sufficiently small constant (we will show that any $\eta \leq 1/16$ suffices). Using the information collected in the previous period, node $v$ computes a set of *free slots* $F_v$. A free slot $s \in F_v$ is one where no beep was heard in the $b_v + 2$ slots preceeding it, and the $b_v + 1$ following it. An uncolored node $v$ selects a slot $p_v$ to beep uniformly at random from the set of free slots $F_v$. If after beeping node $v$ determines no other node is in the interval $[p_v - b_v, p_v]$ it becomes colored.

Two neighboring nodes are *colliding* if they beep at the same slot. Every period nodes select independently at random a jitter which affects where they beep. Therefore two collided nodes will detect the collision and become uncolored with constant probability (proof in Appendix B).

**Lemma 5.1.** *If neighboring nodes $u$ and $v$ collide in JITTERANDJUMP, they become uncolored in the next period with probability at least $\frac{1}{2}$.*

By adjusting $\kappa$ and $\eta$ appropriately, it's possible to guarantee that the number of free slots observed by each node is a constant fraction of the number of slots.

**Algorithm 2** JITTERANDJUMP running at node $v$

1: $colored_v \leftarrow false$
2: $S \leftarrow \texttt{listen}[Q]$
3: $\tilde{d}_v \leftarrow \max(|S|, 1)$
4: $b_v \leftarrow \eta \frac{Q}{\tilde{d}_v + 1}$
5: **loop**
6:     **if** not $colored_v$ **then**
7:         $F_v \leftarrow \{p \mid S \cup \{p_v\} \, [p - b_v - 2, p + b_v + 1] = \varnothing\}$
8:         $p_v \leftarrow \mathfrak{U}_{F_v}$
9:     **end if**
10:    $jitter_v \leftarrow \mathfrak{U}[0..1]$
11:    $S \leftarrow \texttt{listen}[p_v + jitter_v - 1] \cup \texttt{beep} \cup \texttt{listen}[Q - p_v - jitter_v]$
12:    $I_v \leftarrow \max s \text{ s.t. } S[p_v - s, p_v] = \varnothing$
13:    $\tilde{d}_v \leftarrow \max(|S|, 1)$
14:    $b_v \leftarrow \eta \frac{Q}{\tilde{d}_v}$
15:    **if** $S[p_v - b_v, p_v + b_v] = \varnothing$ **then**
16:        $colored_v \leftarrow true$
17:    **else if** $S[p_v - 1, p_v + 2] \neq \varnothing$ **then**
18:        $colored_v \leftarrow false$
19:    **end if**
20: **end loop**

**Proposition 5.2.** *If $\kappa \geq 4/\eta$ and $\eta \leq 1/3$ then $|F_v| \geq (1 - 3\eta)Q$ for every node $v$.*

We've established that the degree estimate is an upper bound on the real degree; we also show that with constant probability it is a lower bound on the number of uncolored nodes.

**Lemma 5.3.** *With probability $\frac{1}{2}$ the number of beeps observed by a node is at least a quarter of the number of its uncolored neighbors.*

*Proof.* Fix node $v$ and let $P \subseteq N(v)$ be its uncolored neighbors. We want to show $\mathbb{P}\left[\tilde{d}_v > |P|/4\right] \geq \frac{1}{2}$.

Each node $u \in P$ beeps at random in $F_u$ and if $\kappa \geq 4/\eta$ then from Proposition 5.2 $|F_u| \geq (1 - 3\eta)Q = (1 - 3\eta)\kappa\Delta$. If we let $\eta \leq 1/16$ then $\kappa \geq 1/(1 - 3\eta)$ and thus $|F_u| \geq \Delta$.

On the other hand, the probability of collisions (and a lower degree estimate $\tilde{d}_v$) is increased if $\forall u, w \in P$ $F_u = F_v$. In other words, if $|P| \leq \Delta$ beeps are randomly distributed in $|F_v| \geq \Delta$ slots, and we want to show that the with probability $\frac{1}{2}$ the number of occupied slots is $|P|/4$. This can be cast as a balls and bins problem, where the number of balls is less than the number of bins. Due to lack of space, the balls and bins analysis is presented in Appendix A. □

To argue termination we partition nodes into good and bad nodes. Informally, a good node is one which, modulo the jitter, continues to beep at the same slot in the rest of the execution.

**Definition 1.** *Node $v$ is* good *if it is colored and there does not exist a neighboring node $u \in N(v)$ with a phase $p_u$ such that $|\mathring{p}_u - \mathring{p}_v| \leq 1$; otherwise $v$ is* bad.

By definition, once a node becomes good no neighboring node is colliding with it. Moreover, nodes always listen before beeping and beep at slots which were previously unoccupied. Therefore, it is not surprising that once a node becomes good, it remains good thereafter (proof in Appendix B).

**Lemma 5.4.** *Once a node is* good*, it remains* good *for the rest of the execution.*

We classify bad nodes further as colored and uncolored. First we consider the easier case of colored bad nodes.

**Lemma 5.5.** *A colored* bad *node becomes* good *or uncolored with probability* $\geq \frac{1}{2}$.

*Proof.* Fix a colored bad node $v$. By definition a nonempty set of its neighbors $P \subseteq N(v)$ beep at the same slot as $u$.

If all nodes in $P$ are uncolored then they all jump to a random slot and node $v$ becomes good. Otherwise there exists a colored node $u \in P$. However by Lemma 5.1 with probability $\frac{1}{2}$ in the next period nodes detect the collision become uncolored. □

Now we consider uncolored bad nodes.

**Lemma 5.6.** *An uncolored* bad *node becomes* good *with probability* $\geq \frac{1}{2} e^{-\frac{16\eta}{1-3\eta}}$.

*Proof.* Fix an uncolored bad node $v$. Let $B_u$ be the event that node $u$ choses to beep in the interval $[p_v - b_v, p_v + b_v]$. In other words, $B_u$ is the event that node $u$ interferes with the beep of $v$. By definition $\mathbb{P}[B_u] \leq \frac{2b_v}{|F_u|}$, and from Proposition 5.2 $|F_u| \geq Q(1-3\eta)$ and thus $\mathbb{P}[B_u] \leq \frac{2b_v}{Q(1-3\eta)} \leq \frac{2\eta}{\tilde{d}_v(1-3\eta)}$.

Let $G_v$ be the event that node $v$ becomes good. Node $v$ becomes good unless a non-empty subset of its (uncolored) neighbors choose a random slot that interferres with its beep. Hence $\mathbb{P}[G_v] = \prod_{u \in P} \mathbb{P}[\neg B_u]$ where $P \subseteq N(v)$ are the uncolored neighbors of $v$.

Let $P_v$ be the event that the number of beeps observed by $v$ is at least one quarter of the number of its uncolored neighbors, that is $\tilde{d}_v \geq |P|/4$. We show that conditioned on $P_v$, node $v$ becomes good with constant probability.

$$\mathbb{P}[G_v|P_v] = \prod_{u \in P} \mathbb{P}[\neg B_u|P_v] = \prod_{u \in P}(1 - \mathbb{P}[B_u|P_v]) \geq \left(1 - \frac{8\eta}{|P|(1-3\eta)}\right)^{|P|} \geq e^{-\frac{16\eta}{1-3\eta}}$$

Where the last inequality holds for sufficiently small $\eta \leq \frac{1}{16}$. Finally from Lemma 5.3 we have $\mathbb{P}[P_v] \geq \frac{1}{2}$, hence $\mathbb{P}[G_v] \geq \mathbb{P}[G_v|P_v]\mathbb{P}[P_v] \geq \frac{1}{2} e^{-\frac{16\eta}{1-3\eta}}$. □

From Lemmas 5.5 and 5.5, after two periods a bad node becomes good with constant probability. Therefore the probability that a node remains bad drops off exponentially with the number of periods. Using standard arguments (see Lemma B.1) one can show that a bad node becomes good with high probability after $\frac{6}{e^{-\frac{16\eta}{1-3\eta}}} \log n \in \mathcal{O}(\log n)$ rounds.

Finally, we show that the output is indeed a valid $\Omega(T/d^{\max}(v))$-interval coloring.

**Lemma 5.7.** *Let $v$ be a* good *node, then* $I_v \geq \eta \frac{Q}{2d^{\max}(v)+1}$.

*Proof.* Consider the period when $v$ became colored. By construction node $v$ observed no beeps in the interval $[p_v - b_v, p_v]$, thus $I_v \geq \eta \frac{Q}{\tilde{d}_v+1}$.

Fix a node $u \in N(v)$. Node $u$ will only select to beep in phases that respect a buffer of size $b_u + 2 = \eta \frac{Q}{\tilde{d}_u+1} + 2$ before the beep of node $v$. So independent of the jitter, node $v$ will never observe a beep of $u$ within within $b_u$ of its phase. Finally, since $\forall u \in V$ it holds that $\tilde{d}_u \leq 2d^{\max}(v)$, we obtain $I_v \geq \eta \frac{Q}{2d^{\max}(v)+1}$. □

This leads to our main theorem.

**Theorem 5.8.** *Node $v$ becomes* good *in $\mathcal{O}(\log n)$ periods after waking up with high probability, with an interval of size $\Omega(T/d^{\max}(v))$.*

## 5.1 Dynamic Graphs

Let us now turn our attention to dynamic graphs, where nodes and edges are added and removed throughout the execution. Adding nodes or edges is analogous to waking up, which is already handled gracefully by JITTERANDJUMP; however this is not the case for node or edge removals. In particular, once the algorithm has stabilized to an $\Omega(T/\Delta)$-interval coloring, the interval of each node is not guaranteed to increase, even if sufficiently many nodes leave and the new maximum degree becomes $\Delta' \ll \Delta$.

A natural solution would be to go back to an uncolored state when the degree estimate falls below a certain threshold. However, colliding nodes can cause the degree estimate to drop artificially, even when no nodes or edges are removed. In some cases, the colliding nodes are not aware of each other and can remain collided forever despite jittering. For example in a star graph, from the center's perspective the spokes may be colliding, but they have no means of detecting the collision.

*Algorithm description (modifications to JITTERANDJUMP).* Regardless of the state, each node $v$ picks a second phase $p'_v$ at random from the free slots $F_v$. In addition to beeping at $p_v + jitter_v$ as before, node $v$ will also beep at $p'_v$. Let $S_v(i)$ be the set of slots where node $v$ heard a beep in period $i$. We define $d^*_v(i) = \max_{j \in [i-r,i]} |S_v(j)|$ as the maximum number of beeps over a moving window of the last $r$ periods. At period $i$ we update the degree estimate by taking the maximum of the current beep count and $d^*_v(i)$ ( $\tilde{d}_v = \max(\tilde{d}_v, d^*_v(i))$ ). Finally, if $d^*_v(i) < \frac{\tilde{d}_v}{16}$ we set $\tilde{d}_v = d^*_v(i)$ and uncolor node $v$.

Since nodes beep twice at every period then $S_v(i) \leq 4d(v)$. In executions where the degree estimate doesn't decrease, the analysis of Section 5 holds with slightly different constants. To prove correctness we need to show that with sufficiently high probability the degree estimate will decreases if and only if the degree drops by a large enough factor.

From proposition 5.2 the number of free slots is $|F_v| \geq (1-3\eta)Q = (1-3\eta)\kappa\Delta$, and since $\kappa \geq \frac{1}{1-3\eta}$ then $|F_v| \geq \Delta$. Given that a node $v$ has $d(v)$ neighbors, and each neighbor beeps at least once per period in a random slot (at most twice), we are interested in the probability that the beeps observed account for a constant fraction of the neighbors. This is essentially the same scenario described by lemma 5.3 which can be viewed as an occupancy problem (see appendix A). We can show that with probability at least $\frac{1}{2}$ the number of beeps observed is at least $d(v)/4$.

Hence, with probability $\geq \frac{1}{2}$ at every period $|S_v(i)| \geq d(v)/4$. Since the degree estimate is computed using the information of the last $r$ periods, the degree estimate decreases only if in the last $r$ periods the beep count observed was below $\tilde{d}_v/16$. However, unless the real degree has decreased by a constant factor, this happens with probability less than $\frac{1}{2}^r$. On the other hand, if the real degree decreases by a large enough factor, the degree count observed for the next $r$ periods will be at most four times the real degree, which will cause the degree estimate to decrease with certainty after $r$ periods.

By setting $r \in \mathcal{O}(\log 1/\varepsilon)$ the same argument used before (see lemma B.1) can be used to prove the algorithm described computes an $\Omega(T/\Delta)$-interval coloring in $\mathcal{O}(\log 1/\varepsilon)$ periods with probability $1 - \varepsilon$.

## 6 Lower Bound

We consider a stronger model, namely standard synchronous local broadcast with messages of constant size. During each slot a node sends a message of constant size and receives the set of messages sent by its neighbors. Assume every node $v$ knows its own degree $d(v)$, the maximum degree $\Delta$ and the size of the network $n$, but does not have unique IDs. All nodes start the execution (wakeup) simultaneously.

The rest of this section is devoted to proving the following theorem.

**Theorem 6.1.** *Under the model described, it is not possible to compute an $\Omega(T/\Delta)$-interval coloring or a $\mathcal{O}(\Delta)$ vertex coloring with high probability in less than $\mathcal{O}(\log n)$ slots.*

*Proof.* Let $G_i = (B_i, E_i)$ be a graph on four vertices, with vertex set $B_i = \{a_i, b_i, c_i, d_i\}$ and edge set $E_i = \{(a_i, b_i), (b_i, c_i), (c_i, d_i), (a_i, c_i), (b_i, d_i)\}$. Define $G$ as the graph the cycle graph generated by pasting together $n/4$ copies of $G_i$, where $\forall i \in [\frac{n}{4}]$ the component $G_i$ is connected with the component $G_{i+1 \mod \frac{n}{4}}$ with the edge $(d_i, a_{(i+1 \mod \frac{n}{4})})$. $G$ is a 4-regular graph of size $n$ and inside every component $G_i$ the vertices $b_i$ and $c_i$ have the same closed neighborhood (see Figure 2 in appendix).

Let $s_u^k$ be the state of node $u$ at slot $k$, and let $m_u^k$ be the message sent by node $u$ in slot $k$. Regardless of its state, a node can only choose to send a message amongst a set of constant size of possible messages, let $c$ be the size of this set.

Consider a component $B_i$, and assume the states of $b_i$ and $c_i$ are identical at slot $k$. Since their closed neighborhood is identical, if they send the same message at slot $k$, they will receive the same set of messages and remain in identical states at slot $k+1$. Formally, if $s_{b_i}^k = s_{c_i}^k$ and $m_{b_i}^k = m_{c_i}^k$ then $s_{b_i}^{k+1} = s_{c_i}^{k+1}$.

Moreover, if $b_i$ and $c_i$ are in the same state at slot $k$, they choose what to send according to the same probability distribution, in particular let $p_i$ (where $i \in [1, c]$) be the probability of sending the $i^{\text{th}}$ message. By definition $\sum_{i=1}^{c} p_i = 1$, and thus by Cauchy-Schwarz we have $\sum_{i=1}^{c} p_i^2 \geq \frac{1}{c}$

We prove a lower bound on the probability that $b_i$ and $c_i$ remain in the same state in the next slot:

$$\mathbb{P}\left[s_{b_i}^{k+1} = s_{c_i}^{k+1} \mid s_{b_i}^k = s_{c_i}^k\right] \geq \mathbb{P}\left[m_{b_i}^k = m_{c_i}^k \mid s_{b_i}^k = s_{c_i}^k\right] = \sum_{i=1}^{c} p_i^2 \geq \frac{1}{c}$$

Therefore, if nodes $b_i$ and $c_i$ start at the same state ($s_{b_i}^0 = s_{c_i}^0$) the probability that they remain in the same state after $\ell$ slots is $\mathbb{P}\left[s_{b_i}^\ell = s_{c_i}^\ell \mid s_{b_i}^0 = s_{c_i}^0\right] \geq \frac{1}{c}^\ell$. If we let $\ell = \log_c \frac{n}{4}$ then $\mathbb{P}\left[s_{b_i}^\ell = s_{c_i}^\ell \mid s_{b_i}^0 = s_{c_i}^0\right] \geq \frac{4}{n}$, and thus $\mathbb{P}\left[s_{b_i}^\ell \neq s_{c_i}^\ell \mid s_{b_i}^0 = s_{c_i}^0\right] \leq 1 - \frac{4}{n}$.

Since there are no unique identifiers, initially all nodes have the same state $\forall u, v \in V \; s_u^0 = s_v^0$), and the probability that after $\ell$ slots every component $B_i$ has $s_{b_i}^\ell \neq s_{c_i}^\ell$ is:

$$\mathbb{P}\left[\forall B_i, s_{b_i}^\ell \neq s_{c_i}^\ell\right] = \prod_{i=1}^{n/4} \mathbb{P}\left[s_{b_i}^\ell \neq s_{c_i}^\ell\right] \leq \left(1 - \frac{4}{n}\right)^{\frac{n}{4}} \leq \frac{1}{e}$$

Therefore there exists a pair of neighboring nodes that remain in the same state after $\ell$ slots with constant probability.

$$\mathbb{P}\left[\exists (u,v) \in E \text{ s.t. } s_u^\ell = s_v^\ell\right] \geq \mathbb{P}\left[\exists B_i \text{ s.t. } s_{b_i}^\ell = s_{c_i}^\ell\right] = 1 - \mathbb{P}\left[\forall B_i, s_{b_i}^\ell \neq s_{c_i}^\ell\right] \geq 1 - \frac{1}{e}$$

Moreover, since $G$ is a 4-regular graph, it should ensure interval lengths of size $\Omega(Q/4) \in \Omega(Q)$. Finally, if two nodes in the same state select intervals of size $\Omega(Q)$ slots out of a total of $Q$ slots, the probability that they select overlapping intervals is greater than a constant (or in general $\mathcal{O}(1/\Delta)$). Therefore with constant probability after $\Omega(\log n)$ slots there is a pair of neighboring nodes which do not have an $\mathcal{O}(Q/\Delta)$ coloring.

Observe that if instead of solving interval coloring we were considering vertex coloring, the probability that two neighboring nodes select the same color out of $\Delta$ available colors is also a constant, and thus with constant probability a pair of neighboring nodes select the same color. Which concludes the proof. $\square$

In light of the upper bound of $\mathcal{O}(\log n)$ periods presented in Section 5, the previous bound is asymptotically tight for constant degree graphs. Since each period has $Q \in \mathcal{O}(\Delta)$ slots this implies a lower bound of $\Omega(\log n/\Delta)$ periods for general graphs. If we additionally assume each node beeps at most $\mathcal{O}(1)$ times per period, the same argument yields a lower bound of $\Omega(\log n/\log \Delta)$ periods for general graphs, since for each node the probability of beeping in the same slot as a neighbor is $1/\kappa\Delta$.

# A  Balls and Bins

Consider the problem of placing $m$ balls randomly into $n$ bins, by putting each ball into a bin selected independently at random. We focus on the case where there is at least one bin per ball ($n \geq m$).

Let $A_i^j$ be the event that the $j^{\text{th}}$ ball is placed in the $i^{\text{th}}$ bin. Since each ball picks a bin uniformly at random, then $\mathbb{P}\left[A_i^j\right] = \frac{1}{n}$. Define $Z_i$ as the random variable that describes the number of balls placed into the $i^{\text{th}}$ bin. We want to lower bound the probability that bin $i$ is occupied (contains at least 1 ball).

$$
\begin{aligned}
\mathbb{P}\left[Z_i > 0\right] &= \mathbb{P}\left[\bigcup_{j=1}^{m} A_i^j\right] \\
&\geq \sum_{j=1}^{m} \mathbb{P}\left[A_i^j\right] - \sum_{j<j'} \mathbb{P}\left[A_i^j \cap A_i^{j'}\right] = \frac{m}{n} - \sum_{j<j'} \mathbb{P}\left[A_i^j\right] \mathbb{P}\left[A_i^{j'}\right] \\
&= \frac{m}{n} - \binom{m}{2}\frac{1}{n^2} = \frac{m}{n} - \frac{m}{2}\left(\frac{m-1}{n^2}\right) = \frac{m}{n}\left(1 - \frac{1}{2}\frac{(m-1)}{n}\right) \\
&> \frac{m}{2n}
\end{aligned}
$$

We define the indicator variable $I_{Z_i} = \begin{cases} 1 & Z_i > 0 \\ 0 & \text{otherwise} \end{cases}$.

Let $Z$ be the number of occupied bins, by definition $Z = \sum_{i=1}^{n} I_{Z_i}$, thus we can now lower bound the expected number of occupied bins by a constant fraction of the balls.

$$
\mathbb{E}\left[Z\right] = \mathbb{E}\left[\sum_{i=1}^{n} I_{Z_i}\right] = \sum_{i=1}^{n} \mathbb{E}\left[I_{Z_i}\right] = n \cdot \mathbb{P}\left[I_{Z_i} = 1\right] > n\frac{m}{2n} = \frac{m}{2} \tag{1}
$$

For $i \in [1, m]$ let $X_i$ be the placement of the $i^{\text{th}}$ ball, we can define the number of occupied bins as $Z = f(X_1, \ldots, X_m)$. Hence, consider the Doob Martingale sequence $Y_i = \mathbb{E}\left[Z | X_1, \ldots, X_i\right]$ where by definition $Y_0 = \mathbb{E}\left[Z\right]$ and $Y_m = Z$. The placement of one ball changes the expected number of occupied bins by at most one, thus $\forall i \in [1, m]$ it holds that $|Y_i - Y_{i-1}| \leq 1$.

If the sequence $W_1, \ldots, W_k$ is a martingale, where $|W_i - W_{i-1}| \leq c$ for $i \in [1, k]$, then Azuma's inequality states that $\mathbb{P}\left[|W_m - W_0| \geq \lambda\right] \leq 2e^{-\lambda^2/2kc^2}$. Applying this to our context, we can show that with constant probability the number of occupied bins does not deviate from its expectation by more than a constant fraction.

$$
\mathbb{P}\left[|Z - \mathbb{E}\left[Z\right]| \geq \frac{1}{2}\mathbb{E}\left[Z\right]\right] \leq 2e^{-\frac{m}{8}}
$$

Which proves the following theorem:

**Theorem A.1.** *When placing $m$ balls randomly into $n$ bins, if the number of bins is at least the number of balls ($n \geq m$) and the number of balls is large enough ($m \geq 12$), then with probability more than $\frac{1}{2}$ the number of occupied bins is greater than one fourth the number of balls ($\frac{m}{4}$).*

# B  Deferred Proofs

*Proof of Lemma 4.1.* Let $P$ be the set of distinct phases at which node $v$ hears a beep during any execution (possibly infinite). By construction nodes beep at most once every period and once a node starts beeping it beeps at the same phase for all subsequent periods. Hence, even in an infinite execution, it follows that $|P| = m \leq d(v)$, let $P = \{p_1, \ldots, p_m\}$ be the set of phases in ascending order.

Suppose by contradiction that node $v$ remains in the searching state for more than one period; hence $\forall p \in [0, T]$ there exists a beep in $P[p - b_v, p + b_v]$, otherwise the searching state would have stopped before the period ended. If there exists an $i \in [1..m - 1]$ such that $p_{i+1} - p_i \geq 2b_v$ then there would exist no beep in $P[p^* - b_v, p^* + b_v]$ where $p^* = \frac{1}{2}(p_{i+1} + p_i)$. Hence we assume $\forall i \in [1..m - 1]\ p_{i+1} - p_i < 2b_v$ which implies $p_m - p_1 < (m - 1)2b_v$. Similarly if $T - p_m + p_1 \geq 2b_v$ then there would exist no beep in $P[p^* - b_v, p^* + b_v]$ where $p^* = \frac{T + p_m + p_1}{2} \mod T$. Hence we assume $T - p_m + p_1 < 2b_v$ which implies $T - 2b_v < p_m - p_1$. Thus we have:

$$T - 2b_v < p_m - p_1 < (m-1)2b_v$$
$$T < 2mb_v \leq 2m\frac{T}{2(d(v)+1)} = T\frac{m}{d(v)+1} \leq T\frac{d(v)}{d(v)+1} < T \quad \text{– a contradiction.}$$

$\square$

*Proof of Lemma 4.3.* By definition $\varepsilon \geq \varepsilon_v$ and $d^{\max}(v) \geq d(v)$, thus clearly $b_v \geq I_v$ and in fact $\forall u \in N(v)\ b_v \geq I_u$. From Proposition 4.2 we can assume that $\mathring{p}_v \neq \mathring{p}_u$, we proceed in cases:

- $\mathring{p}_u > \mathring{p}_v$. Then node $u$ hears a beep from $v$ at phase $\mathring{p}_v$ before leaving the search state. By construction node $v$ doesn't leave the search state if $\mathring{p}_v \in [\mathring{p}_u - b_v, \mathring{p}_u + b_v]$, and since $b_v > I_v$ this means $\mathring{p}_v \notin [\mathring{p}_u - I_v, \mathring{p}_u + I_v]$ or equivalently $\mathring{p}_u \notin [\mathring{p}_v - I_v, \mathring{p}_v + I_v]$.

- $\mathring{p}_v < \mathring{p}_u$. Then node $v$ hears a beep from $v$ at phase $\mathring{p}_u$ before it leaves the search state. By construction node $u$ doesn't leave the search state if $\mathring{p}_u \in [\mathring{p}_v - b_u, \mathring{p}_v + b_u]$ and since $b_u > I_v$ this means $\mathring{p}_u \notin [\mathring{p}_v - I_v, \mathring{p}_v + I_v]$.

$\square$

*Proof of Lemma 5.1.* Since nodes $u$ and $v$ beep at the same slot then $x = \mathring{p}_u + jitter_u = \mathring{p}_v + jitter_v$, and since $jitter_u, jitter_v \in \{0, 1\}$ there are four possibilities.

| $\mathring{p}_u$ | $jitter_u$ | $\mathring{p}_v$ | $jitter_v$ |
|---|---|---|---|
| $x$ | 0 | $x$ | 0 |
| $x-1$ | 1 | $x-1$ | 1 |
| $x$ | 0 | $x-1$ | 1 |
| $x-1$ | 1 | $x$ | 0 |

In the first two cases nodes beep at the same slot on the next period iff $jitter_u = jitter_v$, which happens with probability $\frac{1}{2}$. On the remaining cases nodes beep at the same slot if both choose the same jitter on the next period, which happens with probability $\frac{1}{4}$ and thus beep at different slots with probability $\frac{3}{4}$. Hence in all cases nodes beep at different slots in the next period with probability $\geq \frac{1}{2}$. The condition in line 17 ensures that if $u$ and $v$ beep at different slots in the next period, they both become uncolored. $\square$

*Proof of Lemma 5.4.* By definition `good` nodes are colored, consider the last period when $u$ became colored. A neighbor $v$ of $u$ whose beep did not collide with $u$'s will respect a buffer around the beep of $u$, and independent of the jitter $|\mathring{p}_u - \mathring{p}_v| \geq b_v$ for all future periods.

Since node $u$ eventually becomes good, then all its collided neighbors $v \in P$ select a random phase $p_v$, and since when selecting a phase nodes respect a buffer around existing beeps (including their own), it follows that $|\mathring{p}_u - \mathring{p}_v| \geq b_v$. Hence, by letting $k \geq 2/\eta$ we ensure $b_v \geq 1$ and the lemma follows. □

**Lemma B.1.** *If a* bad *node becomes* good *with probability $p$ after $c$ periods, then after $\frac{c(q+1)}{p} \log n$ periods all nodes become* good *with probability $1 - \frac{1}{n^q}$.*

*Proof.* Let $X_v$ be the number of periods before node $v$ becomes good, we lower bound the probability that node $v$ remains good for $\frac{c(q+1)}{p} \log n$ periods.

$$\mathbb{P}\left[X_v > \frac{c(q+1)}{p} \log n\right] = \prod_{i=0}^{\frac{(q+1)}{p} \log n - 1} \mathbb{P}[X_v > i | X_v > i - c] \geq (1-p)^{\frac{(q+1)}{p} \log n} \leq e^{-(q+1)\log n} = \frac{1}{n^{q+1}}$$

And by the union bound,

$$\mathbb{P}\left[\exists v \in V \text{ s.t. } X_v > \frac{c(q+1)}{p} \log n\right] \leq \sum_{v \in V} \mathbb{P}\left[X_v > \frac{c(1+1)}{p} \log n\right] \leq \frac{1}{n^q}$$

□

# C  Figures

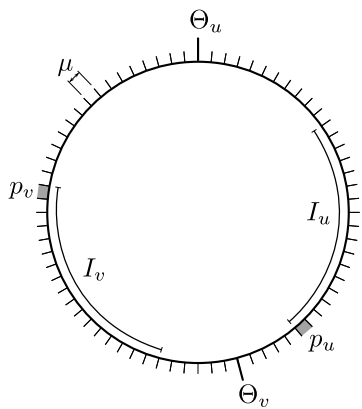

Figure 1: Discrete Interval Coloring Output

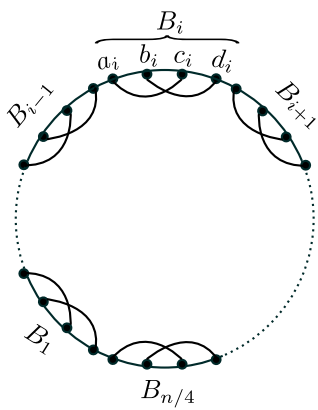

Figure 2: Lower bound graph $G$ on $n$ vertices.